\documentclass[a4paper,11pt]{article}

\usepackage{amsmath,amsfonts,amssymb,caption,graphicx}
\usepackage{ushort}
\usepackage{makeidx}
\usepackage{float}
\usepackage{accents}
\usepackage{color}
\usepackage{framed}
\usepackage{versions}
\usepackage{xcolor}
\usepackage{emptypage}

\def\hmath$#1${\texorpdfstring{{\rmfamily\textit{#1}}}{#1}}

\usepackage[bookmarks]{hyperref}
\hypersetup{
    colorlinks=true,   	
    linkcolor=red,      
    citecolor = [rgb]{0 0.7 0},   	
    filecolor=magenta, 	
    urlcolor=blue
}

\newcommand{\leqa}{\mbox{$ \;\stackrel{(a)}{\leq}\; $}}

\newcommand{\eqb}{\mbox{$ \;\stackrel{(b)}{=}\; $}}

\newcommand{\IND}{{\mathbb I}}

\newcommand{\Po}{\mbox{\rm Po}} 
\newcommand{\Bin}{\mbox{\rm Bin}}

\newcommand{\Bern}{\mbox{\rm Bern}}

\def\ba{\begin{align}}
\def\ea{\end{align}}
\def\ban{\begin{align*}}
\def\ean{\end{align*}}

\def\be{\begin{eqnarray}}
\def\ee{\end{eqnarray}}
\def\ben{\begin{eqnarray*}}
\def\een{\end{eqnarray*}}

\def\bqq{\begin{equation}}
\def\eqq{\end{equation}}
\def\bqqn{\begin{equation*}}
\def\eqqn{\end{equation*}}

\def\elabel#1{\label{e:#1}}

\def\sq{$\Box$}

\def\qed{\ifmmode\sq\else{\unskip\nobreak\hfil
\penalty50\hskip1em\null\nobreak\hfil\sq
\parfillskip=0pt\finalhyphendemerits=0\endgraf}\fi\par\medbreak}

\newsavebox{\junk}
\savebox{\junk}[1.6mm]{\hbox{$|\!|\!|$}}

\def\til={{\widetilde =}}

\def\clB{{\cal B}}

\def\clG{{\cal G}}

\def\clS{{\cal S}}

 \def\eq#1/{(\ref{#1})}

\newtheorem{theorem}{Theorem}[section]
\newtheorem{corollary}[theorem]{Corollary}
\newtheorem{proposition}[theorem]{Proposition}
\newtheorem{lemma}[theorem]{Lemma}
\newtheorem{definition}[theorem]{Definition}

\def\eq#1/{(\ref{e:#1})}

\newcommand{\beqn}[1]{\notes{#1}%
\begin{eqnarray} \elabel{#1}}

\newcommand{\eeqn}{\end{eqnarray} } 

\newcommand{\beq}[1]{\notes{#1}%
\begin{equation}\elabel{#1}}

\newcommand{\eeq}{\end{equation}} 

\def\bdes{\begin{description}}
\def\edes{\end{description}}

\def\notes#1{}

\definecolor{mag}{rgb}{0.7,0,0.3}
\definecolor{dgreen}{rgb}{0.1,0.5,0.1}
\definecolor{dred}{rgb}{.8,0,0}
\definecolor{gray}{rgb}{.8,.8,.8}
\definecolor{brown}{rgb}{0.6451,0.3706,0.1745}

\begin{document}
\title{
Compression and Symmetry\\ of Small-World Graphs and Structures}

\author
{
        I.\ Kontoyiannis
    \thanks{Statistical Laboratory,
	Centre for Mathematical Sciences,
        University of Cambridge,
	Wilberforce Road,
	Cambridge CB3 0WB, UK.
                Email: 
			{\tt yiannis@maths.cam.ac.uk}.
        }
\and
	Y.H.\ Lim
    	\thanks{
		Department of Engineering,
        	University of Cambridge,
		Trumpington Street, Cambridge CB2 1PZ, U.K.
                Email: {\tt limjohnyh@gmail.com}.
        }
\and
	K. Papakonstantinopoulou
    		\thanks{Department of Informatics,
		Athens University of Economics and Business,
		Patission 76, Athens 10434, Greece.
                Email: {\tt katia@aueb.gr }.
	}
\and
	W.\ Szpankowski
    	\thanks{Department of Computer Science,
	Purdue University,
	305 N. University Street,
	West Lafayette, Indiana, 47907-2107, U.S.A.
                Email: {\tt szpan@purdue.edu}.
His work was supported by NSF Center for Science of Information (CSoI)
Grant CCF-0939370, and in addition by NSF Grants CCF-1524312, CCF-2006440,
and CCF-2007238.
	}
}

\date{\today}

\maketitle

\begin{abstract}
For various purposes and, in particular, in the context of data
compression, a graph can be examined at three levels.
Its structure can be described
as the unlabelled version of the graph;
then the labelling of its
structure can be added; and finally,
given then structure and labelling,
the contents of the labels can be described.
Determining the amount of information present at each level
and quantifying the degree of dependence 
between them requires the study of
symmetry, graph automorphism, entropy, 
and graph compressibility.
In this paper, we focus on a class of small-world graphs.
These are geometric random graphs where
vertices are first connected to their 
nearest neighbours on a circle 
and then pairs of non-neighbours are
connected according to a distance-dependent
probability distribution. 
We establish the degree distribution of this model, 
and use it to prove the model's asymmetry 
in an appropriate range of parameters.
Then we derive the relevant
entropy and structural entropy of these
random graphs, in connection with graph compression.

\end{abstract}



\thispagestyle{empty}
\setcounter{page}{0}


\newpage

\setcounter{page}{1}

\section{Introduction}
\label{s:intro}

Our main aim in this work is 
to develop rigorous results 
on structural properties that are 
fundamental to statistical and information-theoretic
problems involving the information shared between 
the labels and the structure of
a random graph, specifically within the class
of {\it small-world graphs}.
For various statistical and signal processing
tasks and, in particular, in the context of data 
compression, the information present in a graph 
can be examined at three levels.
First, its {\em structure} can be described, 
that is, the unlabelled version of the graph. 
Second, its labelling can be described given
its structure. And third, the actual contents of 
the labels can be described, given the structure 
and the labelling~\cite{tray:arxiv}.
In some problems, for example in recovering 
the node arrival order of dynamic 
networks~\cite{nature2019}, the goal is to first \emph{recover} 
label information by examining a graph structure,
and then to explain the structural properties
(such as symmetry) involved in their analysis.

More formally, the labelled and unlabelled
graph compression problems
can be described as follows. Fix a graph model on the 
collection $\clG(n)$ of all simple, undirected,
labelled graphs on 
$n$ vertices. First, we aim to understand the best
achievable performance of efficiently computable 
\emph{source codes} for this model~\cite{coverthomas}. 
A source code $(C_n, D_n)$ here consists 
of an encoder $C_n$ mapping graphs 
in $\clG(n)$ to finite-length bit strings, 
and of a decoder $D_n$ that inverts $C_n$.
The goal is to make
the (expected) length of the output bit string 
as short as possible.
Of particular interest to us here is the
related problem of the compression of graph {\em structures}.
In this case, the encoder $C_n$ is presented 
with a graph $G_n$ isomorphic to a sample from $\clG(n)$, 
and $D_n(C_n(G_n))$ is only required to be a labelled graph 
isomorphic to $G_n$, so that only the structural 
information is preserved. 
We again seek to characterize efficient source codes
with minimal code lengths. This optimal
compression
performance is characterized by the entropy of the distribution 
on unlabelled graphs induced by the model,
which we call its \emph{structural entropy}.

\medskip

\noindent
{\bf Structural properties.}
Several interesting structural properties
and quantities arise naturally in connection with graph
compression. As we describe next,
determining the structural entropy often
involves computing the size of the automorphism group of a graph, 
as well as the typical number of positive-probability 
\emph{labelled representatives} 
(re-labellings or permutations) of a given structure. 

In general, given a labelled graph $G_n$
generated by some model on $\clG(n)$,
all $n!$ label permutations lead to the
same {\em structure} $S_n:=S_n(G_n)$; {\it however},
not all permutations may be permissible under 
the model, and some permutations may lead to the 
exact same graph.
The latter property is well characterized by 
the automorphism group, ${\rm Aut}(G_n)$, of $G_n$.
When the cardinality of the automorphism group  is one, then the
graph is {\it asymmetric} since every feasible permutation is distinct
(in term of the labelled graph) and gives the same structure. 
In some cases, such as the Erd\H{o}s-R\'enyi (ER) 
model~\cite{bollobas:book} and preferential attachment graphs 
(PAG)~\cite{barabasi99,magnerluczak}, all permutations lead to
the same graph with high probability. In other words, 
these models are invariant under
isomorphism. Furthermore, in the ER model every permutation 
is feasible, unlike under the PAG model.
For PAG graphs, the number of distinct re-labellings can be
computed as the ratio of the number of feasible permutations,
$|\Gamma(G_n)|$,
and the size of the automorphism group, $|{\rm Aut}(G_n)|$. 
As a consequence, the
structural entropy of the unlabelled graph
is a function of $\log |\Gamma(G_n)|/|{\rm Aut}(G_n)|$ 
as well as of the (labelled) graph entropy.
However, when the model is 
not invariant under isomorphism,
we need to actually estimate the {\em conditional entropy}
of the (labelled) graph under a given structure.

\newpage

One such class is the family of small-world 
graphs~\cite{watts-strogatz:98,kleinberg:00} on the circle.
In this paper we focus on the symmetry, entropy
and automorphism properties of graphs 
generated by this model.


\medskip

\noindent
{\bf Contributions.}
We study the small-world model~\cite{watts-strogatz:98,kleinberg:00}
where $n$ vertices are arranged on the circle in increasing order,
and each node is connected to its two nearest neighbours. 
Then different pairs of nodes are (independently) connected
with probability proportional to $1/k^a$, where $k$ is their
distance and $a\in(0,1)$ is fixed parameter;
precise definitions are given in 
Section~\ref{sec-sw}.
Such a model does not satisfy the two properties discussed above:
It is not invariant under isomorphism,
and not every permutation is feasible.

For the small-word model we first compute the mean 
degree of a node (Proposition~\ref{prop:meand}), 
and in Theorem~\ref{thm:swasym} we prove 
that it is asymmetric with high probability.
This allows us to derive very accurate
asymptotic estimates for the graph entropy 
and structural entropy;
these are presented in Theorem~\ref{thm:swH}
and Corollary~\ref{cor:swse}.
Finally, in Theorem~\ref{thm:swSH}
we give a precise upper bound on
the conditional entropy of a small-world 
graph given its structure, a result
which is of independent interest 
from both the combinatorial and
information-theoretic points of view.

\medskip

\noindent
{\bf Prior work.}
%
There is a long history of very 
detailed results on the problem
of determining the fundamental limits of the best
achievable compression performance for sequential data;
see, e.g.,~\cite{strassen:64b,kontoyiannis-97,kontoyiannis-verdu:14}
and the references therein.
However,
the study of the compression problem for 
graph and tree models,
in both the information theory and the computer science literature,
is more 
recent~\cite{kiefferyang,verduabbe,abbe,choi-spa:12,%
anantharam2017,mitzenmacherkdd}.
In 1990, Naor~\cite{naor90} proposed an efficiently computable 
representation for unlabelled graphs 
(answering Tur\'an's~\cite{turan84} 
open question),
and showed that this representation
is optimal up to the second leading term 
of the entropy when all unlabelled graphs 
are equally likely. 
Naor's result is, asymptotically, a special case
of corresponding expansions developed later
in~\cite{choi-spa:12}, where 
general ER graphs were analyzed. 
Further extensions to 
PAG graphs were derived in~\cite{luczak19}.

An approach based on automata,
was used in~\cite{mohri15-automata} 
to design an optimal graph compression scheme.
Recently, the authors of~\cite{venkat} proposed a general 
universal lossless source coding algorithm for graphs,
and there are also a number of heuristic 
methods for real-world graph compression, 
including a grammar-based 
scheme for data structures~\cite{mitzenmacherkdd,graph18,peshkin07}.
Efficient compression algorithms 
were developed in~\cite{basu}, 
leveraging symmetry properties
of graphs arising in connection with deep neural networks.
A comprehensive survey of lossless graph
compression algorithms  can be found in~\cite{besta18}.

There are a number of studies of the compression problem
for trees~\cite{zbyszek18,kieffer09,kiefferyang,turowski18,zhang14}.
For binary, plane-oriented trees, rigorous information-theoretic
results were obtained in~\cite{turowski18},
and a universal,
grammar-based lossless coding scheme 
was proposed in~\cite{zhang14}. 

In the computer science literature, the focus has 
been almost exclusively on algorithmic complexity,
and very little attention seems to have been given
to comparisons with fundamental information-theoretic 
compression measures -- which is the main focus of this paper.
Also, work in both communities has largely been restricted to
labelled graphs, or graphs with strong edge independence assumptions
(with the exception of~\cite{aldous-ross:14,turowski18}).  
As we show, interesting additional complications 
arise when the goal is to compress graphical
{\em structures}.

\medskip

\noindent
{\bf Paper organization.} 
In the next section, after some technical 
preliminaries, we review some known symmetry and structural 
entropy properties of the ER and PAG models. 
The small-world graph model is introduced in Section~\ref{sec-sw},
where its degree distribution is determined and its asymmetry established.
Our main results on the graph entropy and structural entropy of
small-world graphs are stated and proved in Section~\ref{sec-analysis}. 

\section{Preliminaries: Random Graphs and Entropy}
\label{s:randomgraphs}

\subsection{Graphs, structures, labels, and symmetry}

\def\PA{{\rm PA}}
\def\parsec{\par\noindent}
\def\med{\medskip\parsec}

Let $\clG(n)$ denote the class of
all (undirected, simple, labelled) graphs $G=(V,E)$
on $n=|V|$ vertices, where for simplicity
we take $V=\{1,2,\ldots,n\}$ throughout. Let $P_n$
denote a {\em model} for such graphs,
that is, a discrete probability mass
function (PMF) on $\clG(n)$. For example,
under the classical 
Erd\H{o}s-R\'{e}nyi (ER) model~\cite{erdos:59,bollobas:book}
with parameter $p$,
for $G\in\clG(n)$,
$$P_n(G)=p^{|E|}(1-p)^{\binom{n}{2}-|E|},$$
where $|E|$ is the number of edges of 
$G=(V,E)$.
On the other hand, for the preferential attachment 
graphs $\PA(m,n)$ studied, e.g., in~\cite{barabasi99},
where a new node connects to $m$ existing
nodes with probability which is proportional 
to their degree,
the probability $P_n(G)$ does not 
depend only on $|E|$.

Any such model $P_n$ induces a probability
distribution $Q_n$ on structures. Let
$\clS(n)$ denote the class of all unlabelled
graphs of $n$ vertices. Then the induced
probability of a structure $S\in\clS(n)$ is 
the sum of the probabilities of all graphs $G$
with the 
same structure $S$,
\be
Q_n(S)=\sum_{G\in{\rm Iso}(S)}P_n(G),
\label{eq:Qn}
\ee
where ${\rm Iso}(S)\subset\clG(n)$ is the
isomorphism equivalence class consisting  
of all graphs in $\clG(n)$ with structure $S$.

Some standard models $P_n$, such as the simple ER
model and $\PA(m,n)$ graphs~\cite{magnerluczak},
are invariant
under isomorphism, that is, $P_n(G)=P_n(G')$,
whenever there $G$ and $G'$ are both permissible
(i.e., they have nonzero probability under $P_n$)
and there
is an $S$ such that both $G,G'\in{\rm Iso}(S)$.
In such cases we simply have,
$$
Q_n(S)=P_n(G)\cdot|{\rm Iso}(S)|,
$$
where $G$ is any graph in ${\rm Iso}(S)$,
i.e., with structure $S$.
If, in addition, every permutation is permissible 
by a given model
-- as in ER model -- then
the number of graphs isomorphic to a given $G$
is equal to the number of permutations of
the labels, $n!$, divided by the number
of such permutations that lead to exactly
the same graph, namely, the size of the
automorphism group $\rm{Aut}(G)$ of $G$.
Therefore, 
\be
Q_n(S)=P_n(G)\cdot\frac{n!}{|\rm{Aut}(G)|}.
\label{eq:aut}
\ee

More generally, 
in cases like the $\PA(m,n)$ model~\cite{magnerluczak}, 
where not all permutations are permissible,
we have,
\be
Q_n(S)=P_n(G)\cdot\frac{|\Gamma(G)|}{|\rm{Aut}(G)|}.
\label{eq:aut1}
\ee
where $\Gamma(G)$ is the set of permissible permutations.
For example, for $\PA(m,n)$ we know that
$E[\log |\Gamma(G)|]=n \log n -O(n\log \log n)$~\cite{magnerluczak}.
[Throughout the paper, $\log$ denotes the natural logarithm $\log_e$.]
As we will see later, the small-world model considered here
is not invariant under isomorphism,
and not every permutation is permissible.

It is of interest to know how much symmetry a given graph has. In particular,
in some applications one needs to know if a graph is asymmetric, as defined below.
\begin{definition}
A graph $G$ is called {\em asymmetric} if
$|\rm{Aut}(G)|=1$.
\end{definition}

\noindent
It is known that, in appropriate parameter ranges,
the ER~\cite{kim-et-al:02} and PAG~\cite{magnerluczak}
models generate asymmetric graphs with high probability:

\begin{theorem}[ER asymmetry~\cite{kim-et-al:02}]
\label{thm:ksv}
{\rm (i)} For a sequence of random graphs $\{G_n\}$
under the 
ER model
with parameters $\{p_n\}$,
such that, as $n\to\infty$,
$$p_n\gg\frac{\log n}{n}\;\;\;\mbox{and}\;\;\;
1-p_n\gg\frac{\log n}{n},$$
we have, for any $t>0$,
$$\Pr(\mbox{$G_n$ is symmetric})=O(n^{-t}),$$
as $n\to\infty$.
\parsec
{\bf (PAG asymmetry~\cite{magnerluczak})}
{\rm (ii)} For a sequence of random graphs $\{G_n\}$
under the $\PA(m,n)$ model with $m\ge 3$,
we have that, for some $\delta>0$,
$$
\Pr(\mbox{$G_n$ is symmetric})=O(n^{-\delta}),$$
as $n\to\infty$.
\end{theorem}
 
\noindent
One of our main results below will be the development of
a statement analogous to Theorem~\ref{thm:ksv}
for a class of small-world random graphs.

\subsection{Entropy and compressibility}

Detailed asymptotic expansions for
the graph entropy $H(G_n)$ 
under the ER and PAG models
are known as we review below.
First we note, without proof, 
a simple expression for the 
binary entropy function.

\begin{lemma}
\label{lem:h}
As $p\to 0$, the binary entropy function
$h(p)=-p\log p-(1-p)\log(1-p)$ satisfies,
$$h(p)=p\log\Big(\frac{1}{p}\Big)+p-\frac{1}{2}p^2+O(p^3).$$
Moreover, the error term
always satisfies $-(1/2)p^3\leq O(p^3)\leq 0$,
and for $p\leq 1/4$ it also satisfies
$O(p^3)\leq -(1/10)p^3$.
\end{lemma}

\begin{lemma}
\label{lem:ERh}
{\rm (i)} 
{\bf (ER graph entropy)}
For a sequence of ER random graphs $\{G_n\}$ 
with parameters $\{p_n\}$,
$$H(G_n)=\frac{n(n-1)}{2}h(p_n),$$
and if $p_n\to 0$ as $n\to\infty$,
$$H(G_n)
=\frac{n(n-1)}{2}
\left[
p_n\log\Big(\frac{1}{p_n}\Big)+p_n-\frac{1}{2}p_n^2+O(p_n^3)
\right].$$
\parsec
{\rm (ii)} {\bf (PAG graph entropy~\cite{sauerhoff,magnerluczak})} 
For a sequence of $\PA(m,n)$ random graphs $\{G_n\}$,
we have, as $n\to\infty$,
\begin{align}
\label{pag-entropy}
        H(G_n)
        = m n\log n + m\left(\log 2m - 1 - \log m! - A\right) n + o(n),
    \end{align}
    where,
    \begin{align*}
        A = \sum_{d=m}^{\infty} \frac{\log d}{(d+1)(d+2)}. 
    \end{align*}
\end{lemma}

\noindent
{\sc Proof. }
We only sketch the proof for the ER model.
By definition, $G_n\sim P_n$
describes $\binom{n}{2}$
independent $\Bern(p_n)$
random variables,
so, 
$H(G_n)=\binom{n}{2}h(p_n)$
and using Lemma~\ref{lem:h}
gives the result.
The proof of~(\ref{pag-entropy}) can be found in~\cite{magnerluczak}.
\qed

For a random graph $G_n$ with structure $S_n$,
the chain rule for entropy implies that,
\be
H(S_n)=H(G_n)-H(G_n|S_n).
\label{eq:chain}
\ee
Using this identity together with
relation~(\ref{eq:aut}) in combination
with Theorem~\ref{thm:ksv}, 
Choi and Szpankowski~\cite{choi-spa:12} for the ER model 
and Luczak et al.~\cite{magnerluczak} for PAG graphs,
establish the following
asymptotic expansions for the entropy of 
ER and PAG random structures. An analogous expansion
for a class of small-world graphs
is established in this paper.

\begin{theorem}[ER structural entropy]
\label{thm:cs}
{\rm (i)} For a sequence of ER random graphs $\{G_n\}$ 
with parameters $\{p_n\}$ that satisfy,
$$p_n\gg\frac{\log n}{n}\qquad\mbox{and}\qquad 1-p_n\gg\frac{\log n}{n},$$
as $n\to\infty$,
we have, for some $\beta>0$,
$$H(S_n)
=\frac{n(n-1)}{2}h(p_n)-\log n! +O\Big(\frac{\log n}{n^\beta}\Big).$$
\parsec
{\bf (PAG structural entropy)}
{\rm (ii)} 
For a sequence of $\PA(m,n)$ random graphs $\{G_n\}$
with $m\ge 3$ we have,
as $n\to\infty$,
 \begin{align}
        H(S_n)
        = (m-1) n\log n + R_n,
    \end{align}
    where $R_n$ satisfies,
    \begin{align*}
        C n
        \leq |R_n|
        \leq O(n\log \log n),
    \end{align*}
    for some nonzero constant $C=C(m)$.
\end{theorem}

\begin{definition}
The {\em compressibility} of 
a random graph $G_n=(V_n,E_n)$
is measured by the average 
number of bits (or, rather, nats)
per edge used in its best possible description,
that is, $C_n=H(G_n)/E(|E_n|)$.
We say that the sequence of random graphs $\{G_n\}$ is
{\em compressible}, if $C_n=O(1)$. 
\end{definition}

Recent studies indicate that many 
real-world examples of large graphs, 
including web graphs and social media graphs,
are compressible. For an extensive discussion
of compressibility in different models 
see~\cite{chierichetti-etal:13}.

For the ER model we note that each
node has $\Bin(n-1,p_n)$ edges,
which is $\approx\Po(np_n)$ for large
$n$, as long as $p_n=o(1)$; to see this,
recall Theorem~1 of~\cite{barbour-hall:84}.
Also, $E(|E_n|)=n(n-1)p_n/2$, so by Lemma~\ref{lem:ERh}
in this case,
$$C_n=\frac{H(G_n)}{E(|E_n|)}\sim-\log p_n,$$
which is unbounded.
Therefore,
in the above sense, the ER model 
with parameters $p_n=o(1)$
is incompressible.
Similarly, for $\PA(m,n)$ graphs,
we have $C_n \sim \log n$,
for $m\ge 3$. 

\section{A small-world model}
\label{sec-sw}

Here we examine a small-world model,
similar to those introduced 
in~\cite{watts-strogatz:98,kleinberg:00}.
More specifically, it is a Newman-Watts-type model~\cite{newman-watts};
also see~\cite{newman:00,chierichetti-etal:13,aldous-ross:14}.
It is a model of {\em geometric} random graphs,
with high-clustering properties that differentiate
them from ER and PAG models \cite{easley:book,newman:18}.

Consider the vertex set $V=\{1,2,\ldots,n\}$
arranged on the circle, with each vertex 
connected by an edge to its two nearest neighbours.
For each one of the remaining $\binom{n}{2}-n$
pairs of vertices $(u,v)$, we add an edge between 
them with probability $p(|u-v|)$, where 
$|u-v|$ is the discrete distance
on the circle and $p_n(k)=c_n k^{-a}$,
for some $a\in(0,1)$ and  with,
$$
c_n=b_n(1-a)\left(\frac{2}{n}\right)^{1-a},
$$
where 
$\{b_n\}$ is a nondecreasing, unbounded sequence
of positive real numbers, with $b_n=o(n^{1-a})$,
as $n\to\infty$. In all the results and discussion
below we implicitly assume that $n$ is large enough
so that all the $p_n(k)$ are less than one,
which is always possible by the assumptions on $b_n$.

The graph shown below is an example of a small world
graph with $n=16$; nearest neighbour edges are shown
in blue and random edges are green. Note that there
are more edges between nearby nodes and fewer between
distant ones.

\vspace{-0.15in}

\hspace{1.7in}\includegraphics[width=2.0in]{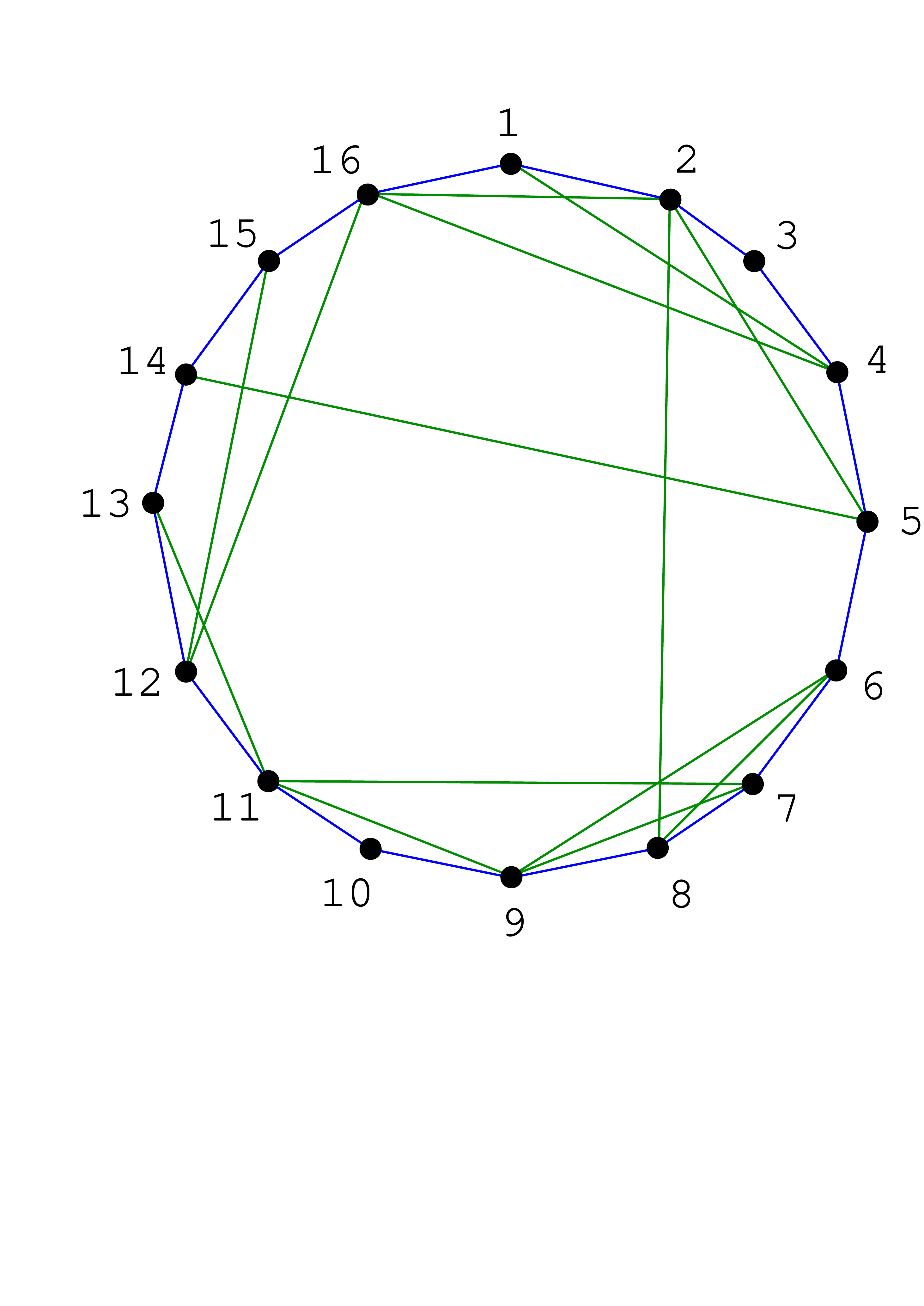}

\vspace{-0.75in}

We call a random undirected graph $G_n$ generated by this model
a {\em random small-world graph with parameters $a$ and $b_n$},
and we write $G_n\sim{\rm SW}(a,b_n)$. It is assumed
throughout that $a\in(0,1)$ and $b_n=o(n^{1-a})$.

Before examining the ${\rm SW}(a,b_n)$ class 
further, we remark that small-world models are an important
class of {\em geometric} random graphs in that, unlike
in the ER model, the connectivity of a small-world graph depends
on the actual locations of the nodes.
Although it will not play a role in our analysis,
we mention that another
important characteristic of such graphs
is the ``small-world property.'' This  means that 
the graph distance between any two nodes is much 
smaller than in a purely random graph, with high 
probability; see the above references or 
the texts~\cite{newman:18,hofstad:16} for details.

\subsection{Degree distribution}

\begin{proposition}[SW mean degree]
\label{prop:meand}
The mean degree $\mu_n$ of an arbitrary node
in a random graph $G_n\sim{\rm SW}(a,b_n)$ satisfies,
$$\mu_n=2b_n+2+O\Big(\frac{b_n}{n^{1-a}}\Big),$$
as $n\to\infty$.
\end{proposition}

\noindent
For the proof we need the following lemmas.
The expansions 
in Lemma~\ref{lem:gamma} follow from 
straightforward applications of Euler-Maclaurin 
summation; see, e.g.,~\cite{apostol:book,apostol:99}.

\begin{lemma}
\label{lem:gamma}
As $n\to\infty$, 
$$
\sum_{k=1}^n\frac{1}{k}
=
	\log n +\gamma+\frac{1}{2n}+O\Big(\frac{1}{n^2}\Big),
$$
where $\gamma$ is Euler's constant
and the error term is bounded in absolute 
value by $\frac{1}{6n^2}$, for all $n\geq 2$.
Also, as $n\to\infty$, 
for any $s>0$, $s\neq 1$, 
$$\sum_{k=1}^n\frac{1}{k^s}
=
	\frac{1}{(1-s)n^{s-1}}
	+\zeta(s) +\frac{1}{2n^s}+
	O\Big(\frac{1}{n^{s+1}}\Big),
$$
where $\zeta$ is the Riemann zeta function,
$$\zeta(s):=
\begin{cases}
\displaystyle
\sum_{k=1}^\infty\frac{1}{k^s},\;\;&\mbox{if}\;s>1,\\
\displaystyle
\lim_{M\to\infty}\left[\sum_{k=1}^M\frac{1}{k^s}
	-\frac{M^{1-s}}{(1-s)}\right],\;\;&\mbox{if}\;s\in(0,1),
\end{cases}
$$
and the error term is bounded in absolute value
by $\frac{s}{6n^{s+1}}$.
\end{lemma}

\noindent
Next we will apply Lemma~\ref{lem:gamma} to get some
simple estimates regarding the probabilities $p_n(k)$.
The proof of Lemma~\ref{lem:Sns} is given in the Appendix.

\begin{lemma}
\label{lem:Sns}
For odd $n$, let:
$$
S_{n,1}=2\sum_{k=2}^{(n-1)/2} p_n(k),
\qquad
S_{n,2}=2\sum_{k=2}^{(n-1)/2} p_n(k)^2.
$$
Similarly, for even $n$, let:
$$
S'_{n,1}=2\sum_{k=2}^{(n-2)/2} p_n(k)+p_n(n/2),
\qquad
S'_{n,2}=2\sum_{k=2}^{(n-2)/2} p_n(k)^2+p_n(n/2)^2.$$
Then, as $n\to\infty$,
$$S_{n,1}=2b_n+O\Big(\frac{b_n}{n^{1-a}}\Big),
\qquad
S_{n,2}=
  \begin{cases}
	\Big(\frac{4(1-a)^2}{1-2a}\Big)
	\frac{b_n^2}{n}
	+O\Big(\frac{b^2_n}{n^{2-2a}}\Big),
   	 	&a\in(0,1/2),\\
	\frac{b_n^2\log n}{n}+O\Big(\frac{b_n^2}{n}\Big),
		&a=1/2,\\
	2^{3-2a}(1-a)^2\zeta(2a)
	\frac{b^2_n}{n^{2-2a}}
	+ O\Big( \frac{b^2_n }{n}\Big),
		&a\in(1/2,1),
  \end{cases}
$$
and the same results hold with $S'_{n,1}$ in place
of $S_{n,1}$, and $S'_{n,2}$ in place of $S_{n,2}$.
\end{lemma}

\noindent
{\sc Proof of Proposition~\ref{prop:meand}. } 
The edges of $G_n$ can be described as 
$\binom{n}{2}-n$ independent Bernoulli random
variables. Choose and fix $n\geq 5$ be arbitrary.

Suppose $n$ is odd. 
Considering, without loss of generality, the node $u=1$,
let
$X_k,Y_k$, for $k=2,3,\ldots,(n-1)/2$,
denote binary random variables,
where each $X_k$ and each $Y_k$ describe
whether node $u=1$ is connected to a different
node at distance $k$ from node $u=1$. Then 
$\{X_k,Y_k\}$ are independent Bernoulli
random variables with corresponding parameters 
$\{p_n(k)\}$,
and the degree of node $u=1$,
$W_n$, say,
can be expressed as $W_n=[2 + \sum_k(X_k+Y_k)]$. 
Therefore, the mean degree of any vertex is,
\be
\mu_n:=E(W_n)=
2+2\sum_{k=2}^{(n-1)/2}p_n(k)=
2+S_{n,1}.
\label{eq:swmd1}
\ee

Similarly, if $n$ is even,
there are
$n$ possible edges between 
pairs of nodes 
at each distance
$k=2,3,\ldots,\frac{n-2}{2}$,
and $n/2$ possible edges
between pairs of nodes at
distance $n/2$. Here,
the mean degree of a vertex
is,
\be
\mu_n=
2+2\sum_{k=2}^{(n-2)/2}p_n(k)+p_n(n/2)
= 2+S'_{n,1}.
\label{eq:swmd2}
\ee
Combining~(\ref{eq:swmd1}) and~(\ref{eq:swmd2})
with Lemma~\ref{lem:Sns}
completes the proof.
\qed

\subsection{Asymmetry}

Let $G=(V,E)$ be an arbitrary undirected graph on $V=\{1,2,\ldots,n\}$,
with no self loops. We first make a series of definitions following
the terminology of~\cite{kim-et-al:02}.

The set of neighbours of a vertex $u\in V$
is denoted,
$$N(u)=\{v\in V\;:\;(u,v)\in E\}.$$
Let $\pi$ be any permutation on $V$. The {\em defect
of a vertex} $u\in V$ under the permutation $\pi$ is,
$$D_\pi(u)=|N(\pi(u))\triangle\pi(N(u))|,$$
which can also be expressed as,
\be
D_\pi(u)=\sum_{v\neq\pi(u)}\left[
\IND{\{(\pi(u),v)\in E,\,(u,\pi^{-1}(v))\not\in E\}}
+\IND{\{(\pi(u),v)\not\in E,\,(u,\pi^{-1}(v))\in E\}}
\right].
\label{eq:Dpi}
\ee
The {\em defect of the graph $G$ under $\pi$}
is,
$D_\pi(G)=\max_{u\in V} D_\pi(u),$
and the {\em total defect of $G$} is,
$$D(G)=\min_{\pi\neq{\sf id}}D_\pi(G),$$
where ${\sf id}$ denotes the identity permutation.
Note that $G$ is asymmetric 
iff $D(G)\neq 0$.

\begin{theorem}[SW asymmetry]
\label{thm:swasym}
$\;$ Let $\{G_n\}$ be a sequence of small-world random graphs,
$G_n\sim{\rm SW}(a,b_n)$, $n\geq 1$. If,
$$b_n=o(n^{1-a}),\qquad\mbox{and}\qquad 
\frac{b_n}{\log n}\to\infty,\qquad\mbox{as}\;n\to\infty,$$
then, for any $t>0$,
$$\Pr(\mbox{\rm $G_n$ is symmetric})=O(n^{-t}),$$
as $n\to\infty$.
\end{theorem}

\noindent
Following~\cite{kim-et-al:02}, 
the proof of 
Theorem~\ref{thm:swasym}, given in the Appendix,
is based in 
part on an application
of the following simple concentration
bound.

\begin{proposition}
\label{prop:conc}
{\em \cite{alon:97,alon-spencer:book}}
Let $Z=f(\xi_1,\xi_2,\ldots,\xi_m)$ be a function
of the independent Bernoulli random variables 
$\{\xi_i\}$,
and suppose that $f$ has the bounded difference
property that, for some $c>0$,
\be
\max_{j,\{\xi_i\}}|
f(\xi_1,\ldots,\xi_{j-1},\xi_j,\xi_{j+1},\ldots,\xi_m)
-f(\xi_1,\ldots,\xi_{j-1},1-\xi_j,\xi_{j+1},\ldots,\xi_m)|\leq c.
\label{eq:bd}
\ee
Let $p_i=E(\xi_i)$ for each $i$, and
$\sigma^2=c^2\sum_{i} p_i(1-p_i)$.
Then, for all $0<t<2\sigma/c$:
$$\Pr\big[|Z - E(Z)|>t\sigma\big]\leq 2e^{-t^2/4}.$$
\end{proposition}

\section{Entropy of the small-world model} 
\label{sec-analysis}

\subsection{Graph entropy}

As with Lemma~\ref{lem:gamma}, 
the expansions in Lemma~\ref{lem:gamma2} below 
are easy applications of 
Euler-Maclaurin summation~\cite{apostol:book,apostol:99}.
It will be used in the proof of 
Theorem~\ref{thm:swH},
given in the Appendix.

\begin{lemma}
\label{lem:gamma2}
As $n\to\infty$, 
$$
\sum_{k=1}^n\frac{\log k}{k}
=
	\frac{1}{2}(\log n)^2 +\gamma'+\frac{1}{2}\frac{\log n}{n}
	+O\Big(\frac{\log n}{n^2}\Big),
$$
where $\gamma'$ is defined, in analogy to Euler's constant,
as,
$$\gamma'=\lim_{n\to\infty}
\left[\sum_{k=1}^n\frac{\log k}{k}-\frac{1}{2}(\log n)^2\right],$$
and the error term is bounded in absolute 
value by $\frac{1+\log n}{6n^2}$, for all $n\geq 2$.
Also, as $n\to\infty$, 
for any $s>0$, $s\neq 1$, 
$$\sum_{k=1}^n\frac{\log k}{k^s}
=
	\frac{\log n}{(1-s)n^{s-1}}
	-\frac{1}{(1-s)^2n^{s-1}}
	-\zeta'(s) +\frac{\log n}{2n^s}+
	O\Big(\frac{\log n}{n^{s+1}}\Big),
$$
where the error term is bounded in absolute value
by $\frac{1+s\log n}{6n^{s+1}}$, for all $n\geq 2$.
\end{lemma}

\begin{theorem}[SW graph entropy]
\label{thm:swH}
Let 
$G_n\sim{\rm SW}(a,b_n)$, $n\geq 1$, 
be a sequence of small-world random graphs
with, 
$$b_n=o(n^{1-a}),\qquad\mbox{and}\qquad 
\frac{b_n}{\log n}\to\infty,\qquad\mbox{as}\;n\to\infty.$$
The entropy of this small-world
model is,
\be
H(G_n)= nb_n \big[\log n-\log b_n -C_a +o(1) \big],
\label{eq:HGn}
\ee
where,
\be
C_a=
	a\Big(\frac{1+\log 2}{1-a}\Big) + 
	\log\Big((1-a)2^{1-a}\Big)
	-1.
\label{eq:Ca}
\ee
\end{theorem}

\noindent
{\bf Remark. }
Note that, combining the above expansion for the
entropy $H(G_n)$ with the expression
for the mean degree of an arbitrary node in $G_n$ given
in Proposition~\ref{prop:meand}, we have that,
as $n\to\infty$,
$$\frac{H(G_n)}{E(|E_n|)}
=\frac{nb_n\log n(\delta_n+o(1))}{2nb_n+O(n)}
\sim\frac{\delta_n}{2}\log n,$$
where the positive sequence $\{\delta_n\}$
is bounded above and bounded away from zero.
Therefore,
the average number of ``bits per edge'' 
in $G_n$ is unbounded, so 
in the terminology of~\cite{chierichetti-etal:13}
the SW$(a,b_n)$ model under our assumptions
is incompressible.

\subsection{Structural entropy}

Having an estimate for the entropy of
a random graph $G_n\sim{\rm SW}(a,b_n)$
in Theorem~\ref{thm:swH}, it is easy
to get a corresponding estimate for the
entropy of the
random structure $S_n$ associated
with $G_n$. Since, given $S_n$,
there are at most $n!\leq n^n$
possible graphs $G_n$ with structure
$S_n$, we have that,
\be
H(G_n|S_n)\leq n\log n.
\label{eq:factorial}
\ee
And since
$H(S_n)=H(G_n)-H(G_n|S_n)$ as noted in~(\ref{eq:chain}),
combining~(\ref{eq:factorial})
with~(\ref{eq:HGn}) immediately 
yields:

\begin{corollary}
[SW structural entropy]
\label{cor:swse}
Under the assumptions of Theorem~\ref{thm:swH},
the entropy of the structures $S_n$
associated with the
small-world random graphs
$G_n\sim{\rm SW}(a,b_n)$
satisfies,
$$H(S_n)= nb_n \big[\log n-\log b_n -C_a +o(1) \big],$$
where the constant $C_a$ is given
in~{\em (\ref{eq:Ca})}.
\end{corollary}

Finally we examine the conditional entropy $H(G_n|S_n)$,
which describes the degree of uncertainty that remains
about the graph $G_n$ after knowing its structure $S_n$.
In Theorem~\ref{thm:swSH} we obtain a slightly 
more refined estimate 
than the crude upper bound in~(\ref{eq:factorial}),
which gives a tighter result when $b_n=o(n^t)$ for all $t>0.$

\begin{theorem}
[SW conditional entropy]
\label{thm:swSH}
Let 
$G_n\sim{\rm SW}(a,b_n)$
be a sequence of small-world random graphs
with associated structures $S_n$,
$n\geq 1$.
Suppose that,
$$b_n=o(n^{1-a}),\qquad\mbox{and}\qquad 
\frac{b_n}{\log n}\to\infty,\qquad\mbox{as}\;n\to\infty.$$
Then the conditional entropy of the graph $G_n$
given its structure $S_n$ has:
$$H(G_n|S_n)\leq
	n\log b_n+(\log 5)n+\log\Big(\frac{n}{b_n}\Big)+O(1).$$

\end{theorem}

First we establish a simple, general upper bound.
As in Section~\ref{s:randomgraphs}, we write $P_n$
for the PMF of $G_n$
on $\clG(n)$ and similarly $Q_n$ for the
induced PMF of $S_n$ in $\clS(n)$.
We also write $\clG_a(n)\subset\clG(n)$ for
the support of $P_n$, and we call graphs
$G\in\clG_a(n)$ {\em admissible}.

\begin{lemma}
\label{lem:tau}
For any graph $G\in\clG_a(n)$ with structure $S$,  let
$\tau(G)$ denote the number of admissible
graphs $G'$ that are isomorphic to $G$, 
$$\tau(G)=|{\rm Iso}(S)\cap\clG_a(n)|.$$
Then:
$$H(G_n|S_n)\leq \sum_{G\in\clG(n)}P_n(G)\log \tau(G).$$
\end{lemma}

\noindent
{\sc Proof. }
First observe that $\tau(G)$ is the same for
all $G\in{\rm Iso}(S)\cap\clG_a(n)$.
Therefore, with only a slight abuse of
notation, we will write $\tau(S)$ for $\tau(G)$
if $S$ is the structure of some $G\in\clG_a(n)$.
In analogy with $\clG_a(n)$, 
let $\clS_a(n)$ denote the set of admissible structures,
i.e., those $S\in\clS(n)$ for which there is an admissible
$G$ with structure $S$.
Then we have,
\ben
H(G_n|S_n)
&=&
	\sum_{S\in\clS_a(n)}Q_n(S)H(G_n|S_n=S)
	\\
&\leqa&
	\sum_{S\in\clS_a(n)}Q_n(S)\log\tau(S)\\
&\eqb&
	\sum_{S\in\clS_a(n)}
	\sum_{G\in{\rm Iso}(S)}P_n(G)\log\tau(S)\\
&=&
	\sum_{G\in\clG(n)}P_n(G)\log\tau(G),
\een
where~$(a)$ follows from the elementary 
fact that the entropy of a random variable with $m$ values
is at most $\log m$, and~$(b)$ follows from the basic 
observation~(\ref{eq:Qn}).
\qed

Next we obtain a simple bound on the tails
of the degree of the nodes of $G_n\sim{\rm SW}(a,b_n)$.
Its proof is given in the Appendix.

\begin{proposition}
\label{prop:tails}
Under the assumptions of Theorem~\ref{thm:swSH},
the probability that there is at least one node in
$G_n$ with degree greater than $9b_n/2$
is $O(n^{-t})$, for any $t>0$.
\end{proposition}

We are now in a position to prove the theorem.

\medskip

\noindent
{\sc Proof of Theorem~\ref{thm:swSH}. }
Let $\clB_n$ be the collection of `bad' graphs $G\in\clG_a(n)$
in the sense of Proposition~\ref{prop:tails},
that have at least one node with degree greater than
$d:=9b_n/2$. For any `good' graph $G\in\clB_n^c$,
we can estimate $\tau(G)$ as follows.
Suppose $G$ has structure $S$ and let $G'\in{\rm Iso}(S)\cap\clG_a(n)$
be not identical to $G$.
Let $\pi\neq{\sf id}$ be the permutation on $V=\{1,2,\ldots,n\}$ that maps $G$
to $G'$. For $G'$ to be admissible it must contain the cycle
of edges $1-2-\cdots-n-1$, which means that $G$ must contain
the cycle,
$$
\pi^{-1}(1)-
\pi^{-1}(2)-\cdots-
\pi^{-1}(n)-
\pi^{-1}(1).$$
So to bound $\tau(G)$ it suffices to
get an upper bound on the number of
permutations $\pi$ with this property.

Fix an arbitrary $i\in V$ as $i=\pi^{-1}(1)$.
Since $G\in\clB_n^c$, the degree of $i$ is at most $d$,
so there are at most $d$ choices for the node $\pi^{-1}(2)$,
and similarly, there are then at most $d$ choices for $\pi^{-1}(3)$.
Continuing this way,
there are at most a total of
$d^{n-1}$ choices for the values of $\pi^{-1}(j)$,
for $j=2,3,\ldots,n$, and an additional $n$ choices for 
the initial value of $i=\pi^{-1}(1)$. Therefore,
there are at most $nd^{n-1}$ possible such permutations,
and so,
$$\tau(G)\leq 
n (5b_n)^{n-1}.$$

Finally, we can substitute this in Lemma~\ref{lem:tau}
to obtain that,
\ben
H(G_n|S_n)
&\leq&
	\sum_{G\in\clB_n^c}P_n(G)\log\tau(G)
	+\sum_{G\in\clB_n}P_n(G)\log\tau(G)\\
&\leq&
	\log\Big[
	n (5b_n)^{n-1}\Big]
	+P_n(\clB_n)\log n!\,,
\een
and using the elementary bound
$n!\leq n^n$, and Proposition~\ref{prop:tails} with
$t=2$, we obtain,
$$H(G_n|S_n)
\leq
	n\log b_n+(\log 5)n+[\log n-\log b_n]+O(1),
$$
as claimed.
\qed


\label{s:conclusions}


\medskip


\section{Conclusions}

This works examines the
degree of compressibility of random graphs
and structures
generated by a one-dimensional version of the 
Newman-Watts
small-world model. 
First,
it is shown that graphs
from that model are asymptotically asymmetric
with high probability,
and the graph entropy of 
the model is computed.
Then, using this symmetry, it is established 
that the structure entropy is 
asymptotically equal
to the graph entropy -- with
equality proved for the
first three (and most significant)
terms in their asymptotic expansion.
Finally, a more accurate bound is given
on the conditional entropy of the random
graph itself given its structure.

Potential applications of this work can
be developed in areas where large graphs 
naturally arise, with characteristics
similar to those in the model examined
here; e.g., 
see~\cite{watts-strogatz:98,newman-watts,kleinberg:00,newman:00,newman:18}
for references to empirical studies involving
graphical data sets with high clustering and
other small-world properties.
In particular, our results can provide
theoretical guidelines for designing
effective compression algorithms for
such data sets, as well as benchmark
values for the fundamental limits 
of the best compression ratios
that can be achieved theoretically.

An interesting and important
direction for future
work is the design of efficient,
near-optimal compression algorithms
for random-world random structures.
These could have important applications
for the communication and storage of
many real-world data sets, 
including, e.g.,
metabolite processing networks, 
neuronal brain networks,
and social influence networks.

Finally we note that 
all the basic estimates in
Lemmas~\ref{lem:h},~\ref{lem:gamma},
and~\ref{lem:gamma2} are given in
nonasymptotic form with closed-form expressions
for the error terms.
Therefore, we expect that 
all the asymptotic results in this paper
can, with some additional work,
be turned into precise,
nonasymptotic, finite-$n$ bounds with explicit
constants. 

\newpage

\appendix

\section*{Appendix: Proofs}

\noindent
{\sc Proof of Lemma~\ref{lem:Sns}. } We only give the proof for odd $n$; 
the case
of even $n$ is similar.

For $S_{n,1}$, by the definition of the $p_n(k)$ we have,
$$S_{n,1}=
2b_n(1-a)\Big(\frac{2}{n}\Big)^{1-a}
\left[\sum_{k=1}^{(n-1)/2}\frac{1}{k^a}-1\right],$$
and by Lemma~\ref{lem:gamma} this is,
$$S_{n,1}=2b_n(1-a)\Big(\frac{2}{n}\Big)^{1-a}
\left[\frac{[(n-1)/2]^{1-a}}{1-a}+O(1)\right]
=2b_n+O\Big(\frac{b_n}{n^{1-a}}\Big).$$

For $S_{n,2}$, we similarly have,
$$S_{n,2}=
2b_n^2(1-a)^2\Big(\frac{2}{n}\Big)^{2-2a}
\left[\sum_{k=1}^{(n-1)/2}\frac{1}{k^{2a}}-1\right],$$
and we apply Lemma~\ref{lem:gamma} in three cases.
For $a\in(0,1/2)$,
\ben
S_{n,2}
&=&
	2b^2_n(1-a)^2\Big(\frac{2}{n}\Big)^{2-2a}
	\left[\frac{[(n-1)/2]^{1-2a}}{1-2a}+O(1)\right]\\
&=&
	\frac{2(1-a)^2}{1-2a}
	b^2_n
	\Big(\frac{2}{n}\Big)
	\Big(\frac{2}{n}\Big)^{1-2a}
	\left[[(n-1)/2]^{1-2a}+O(1)\right]\\
&=&
	\Big(\frac{4(1-a)^2}{1-2a}\Big)
	\frac{b_n^2}{n}
	+O\Big(\frac{b^2_n}{n^{2-2a}}\Big).
\een
For $a=1/2$,
$$S_{n,2}=
\frac{b_n^2}{n}
\left[\sum_{k=1}^{(n-1)/2}\frac{1}{k}-1\right]
=\frac{b_n^2\log n}{n}+O\Big(\frac{b_n^2}{n}\Big).$$
And for $a\in(1/2,1)$,
\ben
S_{n,2}
&=&
	2b^2_n(1-a)^2\Big(\frac{2}{n}\Big)^{2-2a}
	\left[\zeta(2a)+O\Big(\frac{1}{n^{2a-1}}\Big)\right]\\
&=&
	2^{3-2a}(1-a)^2\zeta(2a)
	\frac{b^2_n}{n^{2-2a}}
	+ O\Big( \frac{b^2_n }{n}\Big),
\een
as claimed.
\qed

\noindent
{\sc Proof of Theorem~\ref{thm:swasym}. }
In view of the discussion preceding the theorem,
if 
$G_n\sim{\rm SW}(a,b_n)$,
the probability that it
is symmetric can be bounded above as,
$$\Pr(\mbox{$G_n$ is symmetric})=\Pr(D(G_n)=0)
\leq\sum_{\pi\neq{\sf id}}\Pr(D_\pi(G_n)=0)
=
	\sum_{\pi\neq{\sf id}}\Pr
	\Big(\max_{u\in V}D_\pi(u)=0\Big),$$
and defining, for any $\pi\neq{\sf id}$,
$$Z_\pi=\sum_{u:u\neq \pi(u)}D_\pi(u),$$
we have,
\be
\Pr(\mbox{$G_n$ is symmetric})\leq
	\sum_{\pi\neq{\sf id}}\Pr(Z_\pi=0).
\label{eq:sumxpi}
\ee

To further bound the probability that
$Z_\pi=0$, we will use Proposition~\ref{prop:conc}.
To that end, first observe that,
after ignoring the first term in~(\ref{eq:Dpi}),
we have, for any $\pi$ and any $u$ such that
$u\neq \pi(u)$,
\ben
E[D_\pi(u)]
\geq
\sum_{v\neq u,\pi(u)}
\Pr\big((u,\pi^{-1}(v))\in E,\;
(\pi(u),v)\not\in E\big).
\een
Under the assumptions that $u\neq\pi(u)$ and $v\neq u,\pi(u)$,
the two events in the last probability above 
always refer to two distinct
edges, so they are independent, and hence,
\be
E[D_\pi(u)]
\geq
\sum_{v\neq u,\pi(u)}
\Pr\big((u,\pi^{-1}(v))\in E\big)
\big[1-\Pr\big((\pi(u),v)\in E\big)\big].
\label{eq:reviewer}
\ee
Each term in the last sum
is of the form $p_n(k)[1-p_n(k')]$
for some $k,k'$. 
Therefore, since
$p_n(k)$ is decreasing in $k$ for each $n$,
for odd $n$,
\be
E[D_\pi(u)]
\geq
[1-p_n(2)]
\sum_{v\neq u,\pi(u)}
\Pr\big((u,\pi^{-1}(v))\in E\big)
\geq
[1-p_n(2)]
[S_{n,1}-p_n(2)],
\label{eq:reviewer2}
\ee
with $S_{n,1}$ defined in Lemma~\ref{lem:Sns}.
Note that in the sum that appears 
in~(\ref{eq:reviewer}) and in~(\ref{eq:reviewer2})
we ignore terms that correspond to
edges between nodes at distance $k=1$,
and only sum over pairs at distance $k$ between
$2$ and $(n-1)/2$.
So, by the result of the lemma, we have,
\ben
E[D_\pi(u)]
&\geq&
[1-p_n(2)]\Big[2b_n+O\Big(\frac{b_n}{n^{1-a}}\Big)-p_n(2)\Big]\\
&=&
\Big[1-O\Big(\frac{b_n}{n^{1-a}}\Big)\Big]
\Big[2b_n+O\Big(\frac{b_n}{n^{1-a}}\Big)\Big]\\
&=&
2b_n[1+o(1)],
\een
since $b_n=o(n^{1-a})$. A similar computation
shows that the same result holds for even~$n$.
And letting $d(\pi)$ denote the degree of a permutation $\pi$,
i.e., the number of $u$ such that $\pi(u)\neq u$, we have,
by the above bound and the definition of $Z_\pi$, that:
\be
E(Z_\pi)\geq 2d(\pi)b_n[1 + o(1)].
\label{eq:meanLB}
\ee

Recall that all $D_\pi(u)$ and $Z_\pi$ can be expressed
as functions of the independent 
Bernoulli random variables 
introduced in the proof of Proposition~\ref{prop:meand}.
From the expression in~(\ref{eq:Dpi}) it is clear
that, changing the value of any one of the edges
corresponding to these random variables can only change 
the value of $D_\pi(u)$ by at most~2, and adding or deleting any 
such edge only affects at most four of the terms
in the sum $Z_\pi$. Therefore, 
$Z_\pi$ considered as a function of these
Bernoulli random variables
satisfies the bounded difference property~(\ref{eq:bd}) 
with $c=8$.

For the variance $\sigma^2$ we note that each 
of the $d(\pi)$ many terms
in the sum defining $Z_\pi$ depends
on $(n-3)$ of the corresponding
binary variables.
Therefore, since some of them
may influence $D_\pi(u)$ for more than
one $u$, we can bound, 
for odd $n$,
$$\sigma^2\leq 8^2\times 2\times d(\pi)\times
\sum_{k=2}^{(n-1)/2}p_n(k)[1-p_n(k)]
=\bar{\sigma}^2:=64 d(\pi) [S_{n,1}-S_{n,2}],$$
where $S_{n,2}$ is defined in Lemma~\ref{lem:Sns}.
By the result of the lemma, under the present assumptions
we have $S_{n,2}=o(b_n)$ for all $a\in(0,1)$,
and hence,
\be
\sigma^2\leq 
\bar{\sigma}^2
:=
128 d(\pi)b_n[1+o(1)].
\label{eq:sigmaUB}
\ee

On the other hand, for each $u$ in the definition of $Z_\pi$,
considering the influence on $D_\pi(u)$ of only those $v\neq\pi(u)$ 
that lie on the ``right''
of $\pi(u)$ on the circle (in order to avoid double-counting
edges),
and arguing exactly as above,
we obtain a corresponding lower bound,
\be
\sigma^2\geq 
\underline{\sigma}^2:=
64 d(\pi)b_n[1+o(1)].
\label{eq:sigmaLB}
\ee
Analogous computations show that the bounds~(\ref{eq:sigmaUB})
and~(\ref{eq:sigmaLB}) also hold for even $n$,
and we are now in a position to apply Proposition~\ref{prop:conc}.

Let $N$ be large enough so that, for all $n\geq N$, we have
$120d(\pi)b_n\leq \bar{\sigma}^2\leq 132d(\pi)b_n$ by~(\ref{eq:sigmaUB}),
$\underline{\sigma}^2\geq 60d(\pi)b_n$ by~(\ref{eq:sigmaLB}),
and
the lower bound in~(\ref{eq:meanLB}) is at least $d(\pi)b_n$.
Note that $N$ can be chosen independently of $\pi$, since the
$o(1)$ terms in each of these bounds do not depend on $\pi$.

Let $s=\lambda\bar{\sigma}$, for a fixed
$\lambda\in(0,1/132)$. 
Then, for any $\pi\neq{\sf id}$ and all
$n\geq N$, we have, by the choice of $\lambda$ and
the upper bound on $\bar{\sigma}^2$,
\ben
	\Pr(Z_\pi=0)
&\leq&
	\Pr(Z_\pi<d(\pi)b_n(1-132\lambda))\\
&\leq&
	\Pr(Z_\pi<d(\pi)b_n-\lambda\bar{\sigma}^2)\\
&=&
	\Pr(Z_\pi<d(\pi)b_n-s\bar{\sigma}).
\een
And by the definition of $\bar{\sigma}^2$ and the lower
bound on $E(Z_\pi)$,
\ben
	\Pr(Z_\pi=0)\leq \Pr(Z_\pi<d(\pi)b_n-s\sigma)
\leq
	\Pr(Z_\pi<E(Z_\pi)-s\sigma).
\een
Therefore, by the bound in Proposition~\ref{prop:conc},
we obtain,
\be
	\Pr(Z_\pi=0)
\leq
	\Pr(|Z_\pi-E(Z_\pi)|>s\sigma)
\leq
	 2e^{-s^2/4}
=
	 2e^{-\lambda^2\bar{\sigma}^2/4}
\leq
	2e^{-33\lambda^2d(\pi)b_n},
\label{eq:sumxbound}
\ee
as long as,
$$\lambda<\frac{1}{8}
<\frac{\sqrt{5}}{4\sqrt{11}}
=\frac{\sqrt{60d(\pi)b_n}}{4\sqrt{132 d(\pi)b_n}}
\leq \frac{\underline{\sigma}}{4\bar{\sigma}}
\leq \frac{\sigma}{4\bar{\sigma}},$$
which implies $s<\sigma/4=2\sigma/c$.

Finally, we will sum all the probabilities in~(\ref{eq:sumxbound})
as in~(\ref{eq:sumxpi}). Since there are no more than
$n!/(n-d)!\leq n^d$ permutations that fix $(n-d)$ vertices, we have,
that,
$$
\Pr(\mbox{$G_n$ is symmetric})
\leq
\sum_{\pi\neq{\sf id}}\Pr(Z_\pi=0)
\leq
2\sum_{d=1}^n n^d e^{-33\lambda^2 d b_n}
=
2\sum_{d=1}^n e^{d[\log n-33\lambda^2  b_n]},
$$
and since $b_n/\log n\to\infty$, the right-hand side
above is $O(n^{-t})$, for any $t>0$, as claimed.
\qed

\noindent
{\sc Proof of Theorem~\ref{thm:swH}. } 
In the notation of the proof of~Proposition~\ref{prop:meand},
the edges connecting each node on the circle is
described by a collection of independent Bernoulli
random
variables. Therefore, considering
all $n$ nodes and accounting for double-counting,
when $n$ is odd (the case when $n$ is even is
similar),
\ben
H(G_n)
=
	\frac{n}{2}H(\{X_k,Y_k\})
=
	n\sum_{k=2}^{(n-1)/2} h(p_n(k)).
\een
A weaker version of Lemma~\ref{lem:h} is that,
for small $p$, we have,
$h(p)=p\log(1/p)+p-O(p^2)$, where the error term
is between 0 and $p^2$ for $p<1/2$.
Therefore, taking $n$ large enough so
that all $p_n(k)<1/2$, we have that,
\be
H(G_n)
&=&
	-n\sum_{k=2}^{(n-1)/2} p_n(k)\log p_n(k)
	+n\sum_{k=2}^{(n-1)/2} p_n(k)
	-n\sum_{k=2}^{(n-1)/2} \Delta_{n,k}p_n(k)^2
	\nonumber\\
&=&
	anc_n\sum_{k=2}^{(n-1)/2} \frac{\log k}{k^a}
	-\frac{n}{2} (\log c_n) S_{n,1}
	+\frac{n}{2} S_{n,1}
	-\Delta_n\frac{n}{2} S_{n,2},
	\label{eq:entG1}
\ee
for appropriate constants $\Delta_{n,k},\Delta_n$ in $[0,1]$,
where $S_{n,1}$ and $S_{n,2}$ are defined as in Lemma~\ref{lem:Sns}.

Using Lemma~\ref{lem:gamma2}, 
the first term in~(\ref{eq:entG1}) can be expressed as,
\be
&&
	a(1-a)2^{1-a}n^ab_n\left[
	\frac{\log((n-1)/2)}{(1-a)((n-1)/2)^{a-1}}
	-\frac{1}{(1-a)^2((n-1)/2)^{a-1}}+O(1)
	\right]
	\nonumber\\
&&
	=
	n^ab_n\left[
	a(n-1)^{1-a}\log((n-1)/2)
	-a(n-1)^{1-a}\frac{1}{(1-a)}+O(1)
	\right]
	\nonumber\\
&&
	=
	n^ab_n\left[
	an^{1-a}
	\log n
	-
	an^{1-a}
	\Big(\frac{1+\log 2}{1-a}\Big)+O(1)
	\right]
	\nonumber\\
&&
	=
	ab_n n \log n
	-
	ab_n n 
	\Big(\frac{1+\log 2}{1-a}\Big)
	+o(n).
	\label{eq:H1}
\ee
The sum of the second and third terms 
in~(\ref{eq:entG1}), using Lemma~\ref{lem:Sns}, 
are,
\be
&&
 	-\frac{n}{2}\Big[
	\log\Big((1-a)2^{1-a}\Big)+\log b_n-(1-a)\log n-1\Big]
	\left[2b_n+O\Big(\frac{b_n}{n^{1-a}}\Big)\right]
	\nonumber\\
&&
	=
 	n^ab_n\Big[
	(1-a)n^{1-a}\log n
	+n^{1-a}
	-n^{1-a}\log b_n
	-n^{1-a}\log\Big((1-a)2^{1-a}\Big)
	\Big]
	+o(n\log n)
	\nonumber\\
&&
	=
	(1-a)b_n n\log n
	-b_n n \log b_n
	-b_n n \log\Big((1-a)2^{1-a}\Big)
	+b_n n
	+o(nb_n).
	\label{eq:H2}
\ee
And the last term in~(\ref{eq:entG1}), by
Lemma~\ref{lem:Sns}, is $o(nb_n)$ for all $a\in(0,1)$.
Substituting this together with~(\ref{eq:H1}) and~(\ref{eq:H2})
into~(\ref{eq:entG1}), yields,
$$nb_n \left\{\log n-\log b_n
-\left[
	a\Big(\frac{1+\log 2}{1-a}\Big) + 
	\log\Big((1-a)2^{1-a}\Big)
	-1
\right]
+o(1)
\right\},
$$
as required.
\qed

\noindent
{\sc Proof of Proposition~\ref{prop:tails}. }
We give the proof for odd $n$; the
case of even $n$ is similar.

Let $W_n(i)$ denote the (random) 
degree of node $i$ in $G_n$, so that $W_n(1)=W_n$
as in the proof of Proposition~\ref{prop:meand}.
We will apply Proposition~\ref{prop:conc}
to bound the tails of $W_n$. 
Note that,
$E(W_n)=2+S_{n,1}=2b_n+2+o(1)$, by Lemma~\ref{lem:Sns}.
Also, as a function of the Bernoulli variables
$\{X_k,Y_k\}$ introduced in the proof
of Proposition~Proposition~\ref{prop:meand},
$W_n$ satisfies the assumptions
of Proposition~\ref{prop:conc} with $c=1$. And
in this case,
in the notation of Lemma~\ref{lem:Sns},
the variance $\sigma^2$ is,
$$\sigma^2=2\sum_{k=2}^{(n-1)/2}p_n(k)(1-p_n(k))=S_{n,1}-S_{n,2}
=b_n[2+o(1)].$$
Now consider $N$ large enough such that, 
for all $n\geq N$,
$$2b_n\leq E(W_n)\leq 3b_n, 
\qquad\mbox{and}\qquad
b_n\leq \sigma^2\leq 3b_n.$$
Then by the union bound and symmetry, 
we have that, for any $\lambda\in(0,1/2)$ and $n\geq N$,
\ben
\Pr\Big(\max_{1\leq i\leq n} W_n(i)>2(\lambda+2)b_n\Big)
&\leq&
	n\Pr\big(W_n>2(\lambda+2)b_n\big)\\
&\leq&
	n \Pr\big(W_n>E(W_n)+(2\lambda+1) b_n\big)\\
&\leq&
	n \Pr\big(|W_n-E(W_n)|>(2\lambda+1) b_n \big)\\
&=&
	n \Pr\big(|W_n-E(W_n)|>s\sigma\big),
\een
where we took $s=(2\lambda+1)b_n/\sigma$.
Since $0<s<2\sigma$ for $n\geq N$ by our assumptions,
we can apply Proposition~\ref{prop:conc} with
$\lambda=1/4$ to obtain that,
$$
\Pr\Big(\max_{1\leq i\leq n} W_n(i)>9b_n/2\Big)
\leq
2n\exp(-s^2/4)
=2\exp\Big\{\log n-\frac{9b_n^2}{16\sigma^2}\Big\}
\leq 2\exp\Big\{\log n-\frac{3b_n}{16}\Big\},$$
and since $b_n/\log n\to\infty$ as $n\to\infty$,
the result follows.
\qed



\newpage

\bibliographystyle{plain}


\end{document}